\documentclass[aps,amsmath,amssymb,longbibliography,reprint]{revtex4-1}

\usepackage{subfigure}
\usepackage{graphicx}
\usepackage{epstopdf}
\usepackage{dcolumn}
\usepackage{bm}
\usepackage{verbatim}
\usepackage[colorlinks,linkcolor=blue,urlcolor=blue,citecolor=blue]{hyperref}

\begin{document}

\title{Flexible multi-bunch-length operation for continuous-wave x-ray free-electron lasers}

\author{Zihan Zhu$^{1,2}$}\thanks{Present address: SLAC National Accelerator Laboratory, Menlo Park, CA 94025, USA}

\author{Jiawei Yan$^{4}$}
\author{Hanxiang Yang$^{1,2}$}
\author{Duan Gu$^{3}$}
\author{Bart Faatz$^{3}$}
 \email{faatzbart@sari.ac.cn}
\author{Haixiao Deng$^{3}$}
 \email{denghx@sari.ac.cn}
\author{Qiang Gu$^{3}$}
 \email{guq@sari.ac.cn}
 
\affiliation{%
	$^1$Shanghai Institute of Applied Physics, Chinese Academy of Sciences, Shanghai 201800, China\\ $^2$University of Chinese Academy of Sciences, Beijing 100049, China\\ 
    $^3$Shanghai Advanced Research Institute, Chinese Academy of Sciences, Shanghai 201210, China\\
    $^4$European XFEL, 22869 Schenefeld, Germany.
}%

\date{\today}

\begin{abstract}

The X-ray free-electron lasers (XFELs) are cutting-edge instruments pivotal in a broad range of fields, providing high-power X-ray pulses with durations spanning from femtoseconds to attoseconds. One of the critical challenges in XFEL facilities is the simultaneous accommodation of diverse requirements for XFEL operation modes and photon properties across different undulator lines. This paper proposes a dipole-kicker combination in the bunch compressors to vary the electron bunch length for the continuous-wave XFEL facilities driven by a superconducting linac. This method enables optimization of the electron bunch length on a per-bunch basis, tailored to each specific needs of each undulator. Through start-to-end simulations based on the parameters of the Shanghai high-repetition-rate XFEL and extreme light facility, we demonstrate the feasibility of this technique. The results show its effectiveness in enabling simultaneous operations of self-amplified spontaneous emission and externally seeded FEL across different undulator lines, ensuring optimal electron bunch compression for each undulator line. 
\end{abstract}
\maketitle

\section{Introduction}
X-ray free-electron lasers (XFELs) can generate ultrashort X-ray pulses with high brightness~\cite{huang2007review,huang2021features}, facilitating scientific research in the fields of condensed matter physics, material science, chemistry, and life science~\cite{ostrom2015probing, kang2015crystal,son2011multiwavelength}. Unlike synchrotron radiation facilities, XFELs driven by linear accelerators (linacs) face challenges in simultaneously supporting multiple experimental stations. A key feature that enhances the operational flexibility of XFEL facilities is the configuration of multiple undulators, which allows for the simultaneous conduct of different FEL operation modes or scientific experiments with different requirements for photon properties such as photon energy and pulse duration. 

The electron bunches generated in the linac section are typically distributed by a switchyard section to different undulator lines. Therefore, achieving precise control over the characteristics of each electron bunch is essential to customize the FEL output of the different undulator lines to meet the varying demands of simultaneous experiments. Different methods have been proposed to enable multi-beam-energy operation ~\cite{milne2017swissfel,yan2019multi,zhang2019multienergy} and thus significantly broaden the range of tunable photon energies and increase the flexibility of XFEL facilities. Apart from electron beam energy, the requirements on bunch length also vary for different FEL operation modes or experimental stations. For self-amplified spontaneous emission (SASE) \cite{Kondratenko1980,BONIFACIO1984251}, a higher peak current is normally required to obtain a higher peak power and shorter pulse duration \cite{huang2007review}. However, externally seeded FELs \cite{Yu1991,Stupakov2009,Xiang2009,Deng2013,Feng2014} require electron beams with long and flat current profiles to be less sensitive to fluctuations of arrival time and to have more stable output characteristics of the radiation, especially for the cascading scheme \cite{Yu1997,Liu2013,Allaria2013,RebernikRibic2019,Feng22}. Therefore, there is a growing demand for feasible approaches to simultaneously distributing electron bunches with adjustable bunch lengths into multiple undulator lines. 

In most XFEL facilities, bunch length adjustment is realized by varying the compression in the multi-stage magnetic bunch compressors. This can be achieved by introducing different energy distributions imprinted by RF curvature in the upstream accelerating cavities. The efficiency of adjusting the electromagnetic field is dependent on the time it takes to transfer the RF power into the cavity, that is, RF filling time, which can be estimated as
\begin{equation}
    T_{\mathrm{fill}} = \frac{2Q}{\omega},
    \label{filling}
\end{equation}
where $Q$ is the quality factor ($Q$-factor) of the cavity, and $\omega$ is the RF angular frequency. For normal-conducting X-ray facilities running at around 100~Hz, switching within an RF pulse (within $\mathrm{\mu} s$) has been achieved at SwissFEL and is now used to switch between the two undulator lines Athos and Aramis \cite{milne2017swissfel,paraliev2014high}. They produce two bunches within an RF pulse, each with slightly different RF settings. For a high-repetition-rate XFEL facility based on superconducting (SC) technology, the RF pulse length can range from several hundred microseconds up to continuous wave (CW) and the separation between bunches is normally of the order of one microsecond~\cite{tiedtke2009soft, decking2020mhz, galayda2014lcls, zhu2017sclf}.  However, the RF filling time in these facilities is of the order of one millisecond or more due to the much higher $Q$-factor in the SC accelerating cavities ($\sim 10^{10}$), which makes it impossible to change the RF parameters within shot-to-shot bunch spacing. Hence, alternatives are needed to facilitate shot-to-shot control of bunch length.

In this paper, we proposed a novel scheme to achieve multi-bunch-length operation in a CW XFEL facility driven by SC linac. Instead of adjusting RF parameters to change compression, the trajectory in the bunch compressor can be changed from bunch to bunch by integrating a series of kickers into the bunch compressor chicanes. These kickers add small kicks to some bunches with a high repetition rate, facilitating the desired shot-to-shot control of bunch length for XFEL facilities driven by SC linac. In the following, section~\ref{s1} describes the general idea of bunch compression used in FELs. Several strategies used so far for switching the bunches of different bunch lengths between multiple undulator lines are summarized and analyzed in Section \ref{s2}. Section \ref{s3} shows the principle of the dipole-kicker combinations for adjusting bunch compression. The start-to-end optimization results are presented in Section \ref{s4} to demonstrate the proposed scheme. Section \ref{s5} discusses the research and a conclusion is drawn in Section \ref{s6}.

\section{Principle of bunch compression}
\label{s1}

In an FEL facility, the magnetic chicane is universally introduced to achieve longitudinal compression to generate a short electron bunch with the required current profile. For a typical bunch compressor chicane consisting of 4 rectangular dipole magnets, the bending radius $\rho$ and bending angle $\theta$ of electrons deflected in a magnetic field $B$ can be calculated by
\begin{equation}
	{\rm \rho}=\frac{p}{eB}=\frac{\gamma m_ev}{eB},
    \theta = \arcsin{\frac{L_{\mathrm{dipole}}}{\rho}}, 
 \label{bend}
\end{equation}
where $p$ is the relativistic electron momentum, $m_e$ is the electron rest mass, $v$ is the electron velocity and the $\gamma$ is Lorentz factor.  $L_{\mathrm{dipole}}$ is the length of the dipole magnet

Provided the dipole length with respect to other lengths of the chicane involved is neglected, the difference between the path length of beam through the chicane section and the straight trajectory can be written as
\begin{equation}
    \Delta L=2 {L_\mathrm{drift}}\left(\frac{1}{\cos \theta}-1\right).
\end{equation}
where $L_\mathrm{drift}$ is the drift length between the center of the $1^{\rm st}$ and $2^{\rm nd}$ dipole magnets.

In reality, the bunch experiences nonlinear compression in the dispersive section. If the longitudinal phase space transformation is formulated by the transfer matrix expanded to the third order, the correlation between the electron final longitudinal position $z_\mathrm{f}$ and the initial position $z_\mathrm{i}$ can be given by
\begin{equation}
	z_f=z_i+R_{56}\delta+T_{566}\delta^2+U_{5666}\delta^3+O({\delta}^4),
	\label{BC}
\end{equation}
with momentum deviation $\delta$= $\Delta$$p$/$p$, and where each coefficient can be obtained as:
\begin{equation}
	R_{56}\approx-\theta^2L_\mathrm{drift},
	T_{566}\approx-\frac{3}{2}R_{56},
	U_{5666}\approx2R_{56}.
	\label{$R_{56}$}
\end{equation}

In addition to the nonlinear dispersion terms, the energy chirp introduced by RF curvature in the accelerating modules upstream of the dispersive section also contributes to the nonlinearity in the beam longitudinal modulation. The energy deviation correlated with the longitudinal position along the bunch can be expressed as:
\begin{equation}
	\delta=\delta_0+c_1z_i+c_2{z_i}^2+c_3{z_i}^3+O({z_i}^4),
	\label{eq1}
\end{equation}
where $\delta_0$ is the initial uncorrelated energy spread, $c_1$, $c_2$ and $c_3$ are the first-, second-, and third-order terms of the energy chirp fitted with polynomial, respectively.

Substituting the correlated energy spread $\delta$ that is defined in Eq.~\ref{eq1} into the Eq.~\ref{BC}, the longitudinal position after compression can be expressed as:
\begin{equation}
    \begin{aligned}
	z_f&=(1+c_1R_{56})z_i+(c_2R_{56}+c_1^2T_{566})z_i^2\\
	&+(c_3R_{56}+2c_1c_2T_{566}+c_1^3U_{5666})z_i^3+O(z_i^4).
	\label{BCfinal}
	\end{aligned}
\end{equation}

If we ignore here for convenience the nonlinear dispersion coefficients and nonlinear energy chirp, the Eq.~\ref{BCfinal} becomes the simplest expression as follows:
\begin{equation}
	z_f\approx(1+c_1R_{56})z_i,
	\label{eq2}
\end{equation}
After differentiating at both sides of the equation, Eq.~\ref{eq2} comes to
\begin{equation}
    \begin{aligned}
        \Delta z_f=(&1+c_1R_{56})\Delta z_i = \Delta z_i/C,\\
        &C=\frac{1}{1+c_1R_{56}},
        \label{eq3}
    \end{aligned}
\end{equation}
where $C$ is the approximated compression ratio after the modulation. The method used to change the compression so far has always been adjusting the coefficient $c_1$ in Eq.~\ref{eq3}. Two such methods that are used or have been tested in SC accelerators are described in the next section. Changing the $R_{56}$ has not been considered, because changing the dipoles in the chicane would be much too slow. Nevertheless, a method around this problem is subject of this paper and studied in Sec.~\ref{s3} and the following sections.

\section{RF-switching schemes for SC XFELs}
\label{s2}
As an XFEL operating in CW mode, Shanghai high-repetition-rate XFEL and extreme light facility (SHINE) is under construction in Shanghai, China \cite{zhu2017sclf,Huang2023,Liu2023}. It aims at delivering the x-ray pulses from 0.4 to 25 keV with a repetition rate of up to 1 MHz. At the end of the linac section, the electron bunches are distributed into the three undulator lines referred to as the FEL-I, FEL-II and FEL-III, which will be installed in two of the three tunnels. The operation modes of the FEL-I and FEL-III are SASE and self-seeding, while FEL-II generates XFEL pulses based on SASE and external-seeding FELs. Based on the design of SHINE, the different potential strategies for distribution are analyzed in the following part of this section.

The first SC facility where different compression is used for different FELs is FLASH. It runs in a so-called burst mode, with 500-microsecond long RF pulses filled with bunches separated by one microsecond \cite{faatz2016simultaneous}. This bunch pattern is repeated with a frequency of 10 Hz. The 800-microsecond long bunch trains are split into two parts, separated by about 50 $\mu s$, which is about a factor of 100 slower than the change possible in the normal conducting XFELs such as SwissFEL discussed earlier. Therefore, there is a gap in the bunch train of this duration, allowing time for the flat-top kicker for beam distribution and the RF to change parameters. A fraction of the bunches are delivered to FLASH1 with one set of kicker and RF settings, another fraction to FLASH2 with different settings \cite{ackermann2012simultaneous, ayvazyan2015low}. With this scheme, a change of the longitudinal compression can be achieved, resulting in the control of the final photon pulse duration for both FELs independently. A similar system is used for the European XFEL.

\begin{figure}[h!]
	\centering
	 \includegraphics*[width=\columnwidth]{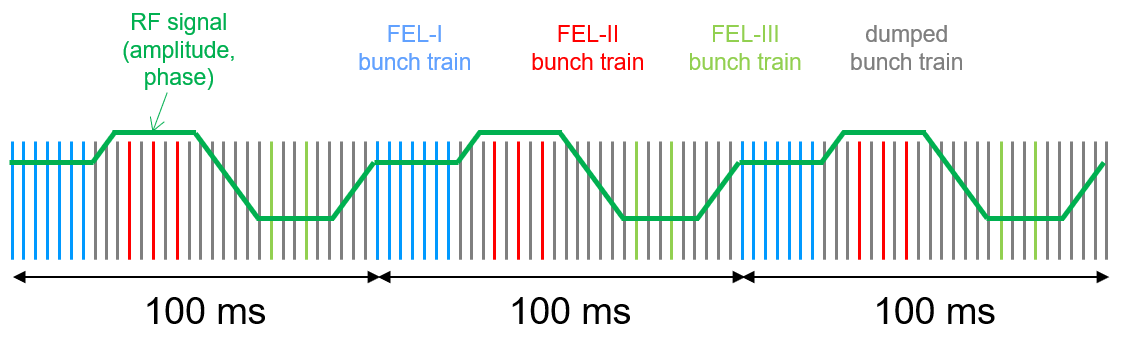}
	\caption{Bunch pattern of burst mode. The bunches between flat-tops and not those needed (both shown in grey) at the end stations are extracted and deflected into the intermediate dump. RF parameters can be customized for each undulator line independently. Each transition needs the order of a ms, which means only a few percent of the total available time is lost.}
	\label{burst}
\end{figure}

For CW hard x-ray XFEL facilities like Linac Coherent Light Source II (LCLS-II) \cite{galayda2014lcls} and SHINE \cite{zhu2017sclf,Huang2023,Liu2023}, the  $Q$-factor of the SC cryomodules is even higher than in FLASH and European XFEL, thus the switching duration is longer than 1 ms. Therefore, the method described above can only be used to deliver short bursts of photon pulses to the experimental stations. To illustrate this bunch operation model, the timing pattern is divided into 10 portions of 100 ms as an example. After 100 ms, the bunch pattern repeats itself. It is shown in Fig.~\ref{burst}. The green line indicates the changes in RF settings. During its flat-top, bunches can be transported to one of the undulators with the required RF settings. During a transition time of up to several ms, RF parameters are changed to deliver optimal conditions for the next undulator, etc. The kicker system for the beam distribution is designed to be fast enough to deliver bunches to an arbitrary undulator on a bunch-by-bunch basis.

\begin{figure}[h!]
	\centering
	 \includegraphics*[width=\columnwidth]{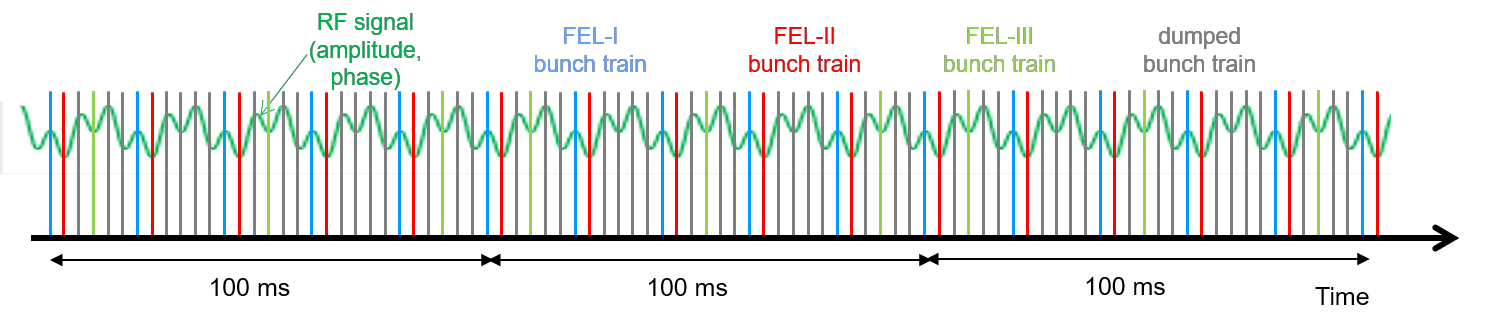}
	\caption{Bunch pattern with detuned cavities, in this case with 3 different frequencies, as shown by the green line. The bunches are distributed equally between undulators. The bunches not needed for each of the undulators are diverted to the intermediate dump and cannot be used by another FEL instead. RF parameters can be customized periodically for each undulator line independently.}
	\label{periodic}
\end{figure}

Another scheme, of which the theory was discussed in Ref.~\cite{zhang2019multienergy} for CW FELs, has been experimentally tested in 2018 at FLASH~\cite{pfeiffer2022rf}. Here, a special cavity is used with a slightly different frequency determined by the minimum bunch spacing. Because of the frequency difference, a modulation of amplitude and phase can be used to set a slightly different amplitude and phase for the two FELs. As a result, bunches can be compressed differently in an interleaved train. Minimally a single cavity with this frequency difference is needed for each bunch compressor. This would be sufficient for FLASH, but for larger differences in compression phases, maybe several cavities will be required. By adding another set of cavities for each compressor with a $3^{\rm rd}$ frequency appropriately chosen, an additional modulation can be used for 3 FELs, etc., as shown in Fig.~\ref{periodic}. Compared to the previous modulation shown in Fig.~\ref{burst}, there is no flat-top RF setting. Instead, a periodic modulation of amplitude and phase is used. In the example shown, a total of 3 frequencies are used, which means that only 3 FELs can have independent settings. This automatically means, that at several locations within the RF pulse, the RF settings cannot be used and the bunch properties at those locations are not suitable for lasing. 

With this scheme, the beam pattern can only be set periodically. If one undulator line is running at 250 kHz and another at 125 kHz, a third beamline cannot use the empty RF buckets with an independent compression phase. The strength of this method is that it is very stable in view of the high $Q$-Factor  of the SC cryomodules. On the other hand, intrinsic stability also means that non-periodic bunch patterns are not possible and many possible bunch locations cannot be used, which becomes worse as the number of independent settings increases due to an increased number of undulators, resulting in a strongly reduced repetition rate. 
An additional disadvantage is that the detuning of standard cavities is limited to about 250~kHz for FLASH. This implies that for two undulator lines, the maximum repetition rate is 250~kHz each. With more undulator lines in parallel, the maximum repetition rate further decreases. For a detuning larger than 250~kHz, dedicated cavities would need to be produced. Also, this amplitude and phase of the linearizing module (at 3.9~GHz) are not modified, which means that the linearization of phase space can only be optimized for one FEL. Therefore, if two FELs use a seeding scheme, which needs a flat current profile, this may become a problem with this scheme.

\section{Changing bunch compressor settings}
\label{s3}

\begin{figure}[h!]
	\centering
	\includegraphics*[width=\columnwidth]{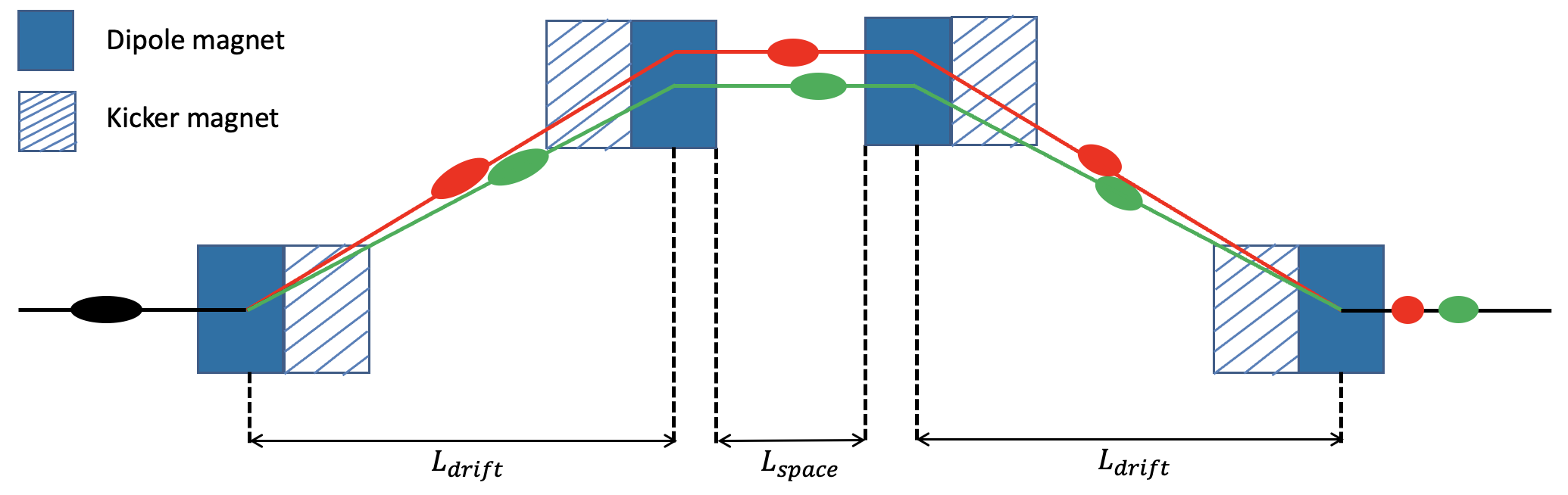}
	\caption{The schematic of dipole-kicker combination in a typical 4-rectangular-dipole chicane.}
	\label{chicane}
\end{figure}
Both approaches described in the previous section focus on adjusting the RF parameters to change the compression for each undulator line. Because the RF filling time is too slow to change the properties between the successive bunches, the parameters $c_1$ cannot be used for fast switching in a CW FEL. Instead, the proposed scheme is based on slightly adapting the dispersive coefficient $R_{56}$. The dipoles themselves are left without any changes during the operation since tuning the magnetic field distribution of dipole magnets within the chicanes is too slow to achieve the modulation on a time scale of microseconds, but a kicker is installed at each dipole within the chicane, as shown in Fig.~\ref{chicane}. 

Changing the $R_{56}$ of the chicane possibly has three consequences. Firstly, the longitudinal dispersion of the chicane is altered. Secondly, the arrival time of the bunch in the RF structures downstream changes. Lastly, there is an impact on the energy gain downstream of the chicane. These impacts are discussed separately below. 


\subsection{Change of $R_{56}$}
As already mentioned in Eq.~\ref{$R_{56}$}, $R_{56}\approx-\theta^2L_\mathrm{drift}$, with $\theta$ the deflection angle by the dipoles. If the deflection angle by the kicker is $\delta\theta$, the kickers alone will create an $R_{56}\approx-(\delta\theta)^2L_\mathrm{drift}$, which is around $10^{-4}$ and will hardly be visible. However, a dipole-kicker combination will add  
\begin{equation}
	\Delta R_{56}\approx-2\theta \delta\theta L_\mathrm{drift},
	\label{eq99}
\end{equation}
 which is of percent level of the main effect and is in practice enough for the small changes needed.

\subsection{Change of effective RF phase}

The deviation in trajectory lengths of the different electron bunch modes is $\Delta L=\Delta R_{56}/2$. This results in the arrival time deviation in the following RF cavities, hence the phase differences of the two modes are expressed as follows:
\begin{equation}
	\Delta \phi_{\rm RF}= 2\pi\times\frac{\Delta L}{\lambda}=2\pi\times\frac{\Delta Lf}{c}\approx 1.56 \,\theta\delta\theta L_\mathrm{drift},
	\label{eq5}
\end{equation}
where $\lambda$ and $f$ are the wavelength and the frequency of the RF pulse, and $c$ is the light speed in the vacuum, where we substituted into the Eq.~\ref{eq5} the L-band (1.3 GHz) cavities. In practice, it means that the change in the compression phase is a fraction of a degree.

\subsection{Change of energy gain}

For a string of cavities, assuming a phase change of $\Delta \phi_{\rm RF}$ as given in Eq.~\ref{eq5}, the total energy change is 
\begin{equation}
	\frac{\Delta E}{E}\approx-\Delta\phi_{\rm RF} (\frac{\Delta\phi_{\rm RF}}{2}+\tan( \phi_{\rm RF}))
	\label{eq98}
\end{equation}
With the original acceleration close to the on-crest phase, Eq.~\ref{eq98} results in a $10^{-4}$-effect, which does not play a significant role, but for a compression phase, the dominant term in Eq.~\ref{eq98} is linear in $\Delta\phi_{\rm RF}$ and will result in a measurable effect. The caused energy discrepancy can not be compensated by adjusting the RF voltage due to the high quality factor in SC XFELs. It is hard to predict if this will result in additional beam loss, especially of halo particles. The optics mismatch, which is a logical consequence, can be compensated after separation of the beams in the distribution area in front of the undulator lines. To avoid accumulation of the effect, for several bunch compressors, the kicker settings have been chosen such, that they add a kick to the dipoles in one compressor, but reduce the kick in the next.
\begin{figure}[h!]
	\centering
	 \includegraphics*[width=\columnwidth]{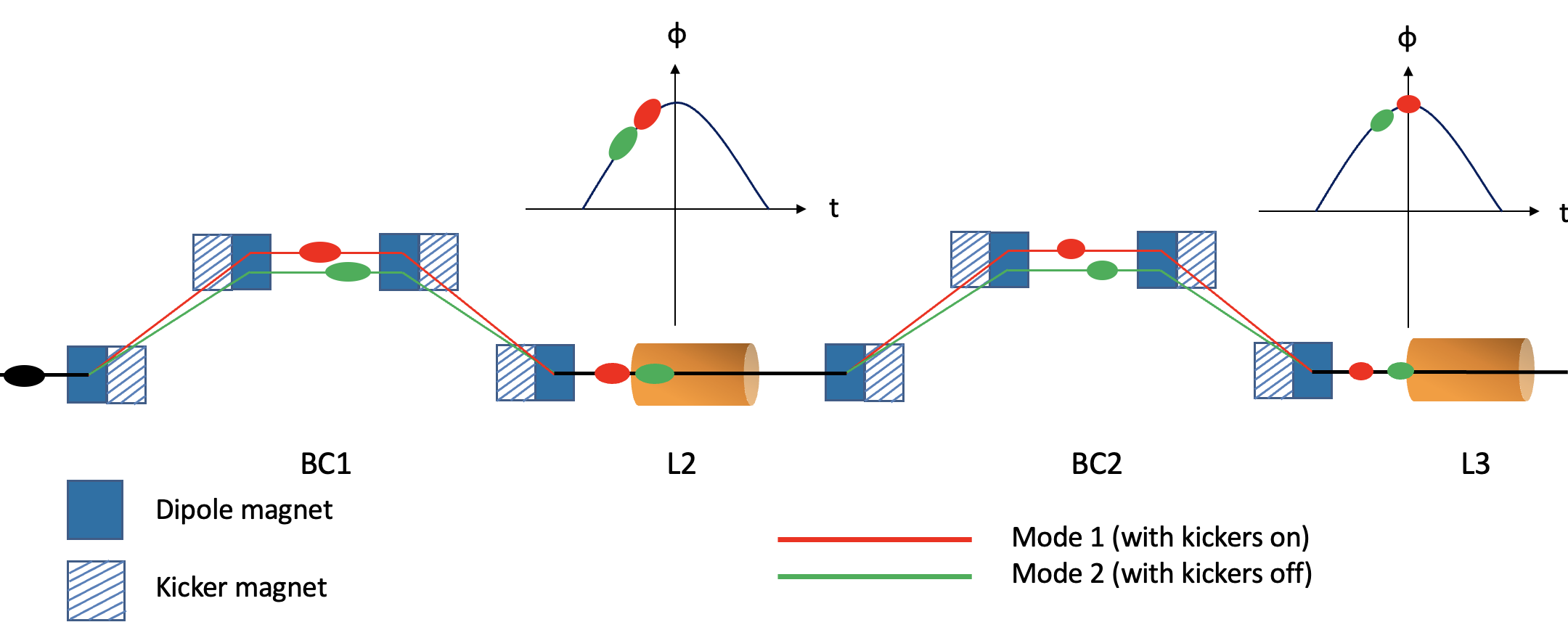}
	\caption{Layout of the typical two-stage bunch compression section in XFELs. The red and green lines indicate the different beam orbits and relative arrival times in the cavities.}
	\label{schematic}
\end{figure}

The schematic of the method for the complete facility is presented in Fig.~\ref{schematic}. The accelerating modules are located between and after the two dispersive sections. The kickers are combined with each individual dipole magnet in the standard 4-dipole C-shape chicane section. The electron bunches travel with the same trajectory until the entrance of the first bunch compressor chicane (BC1) in the linac section. At the first dipole of BC1, one electron bunch mode will travel through the dispersive section with the group of fast kickers powered off (mode 1), while under the influence of the fast kickers, the other bunch will be deflected with a different bending angle $\theta\pm \delta\theta$ and therefore with a different density compression ratio (mode 2). The section is an achromatic structure for both modes with kickers on and off, there is no transverse offset or angle at the exit, only the compression varied. In the following cavity (L2), the bunches from different modes witness slightly different RF phases due to the varied arrival times. Similar to the operation scheme in BC1, in the second magnetic chicane (BC2), the current profile can be managed by slightly changing the $R_{56}$ with the activated fast kickers. As a consequence, the deviation of the traveling path leads to different energy gain in the downstream accelerating cavities.

Additionally, kickers can also be added in the laser heater chicane in the injector section to achieve the arrival time difference in the cryomodules before the bunch compressors. This would cause effectively a phase change in L1, which would be equivalent to a change in $c1$, but because the modules in front of the laser heater chicane are not operated off-crest, the dispersion will not change. The advantage is that the first change is at low energy, which makes the kickers more effective. The disadvantage is that you start in the injector already with an energy mismatch, which needs to be compensated. Therefore, in the following sections of this paper, this option is not pursued.

\section{Simulation Results}
\label{s4}

\begin{figure}[h!]
	\centering
	 \includegraphics*[width=\columnwidth]{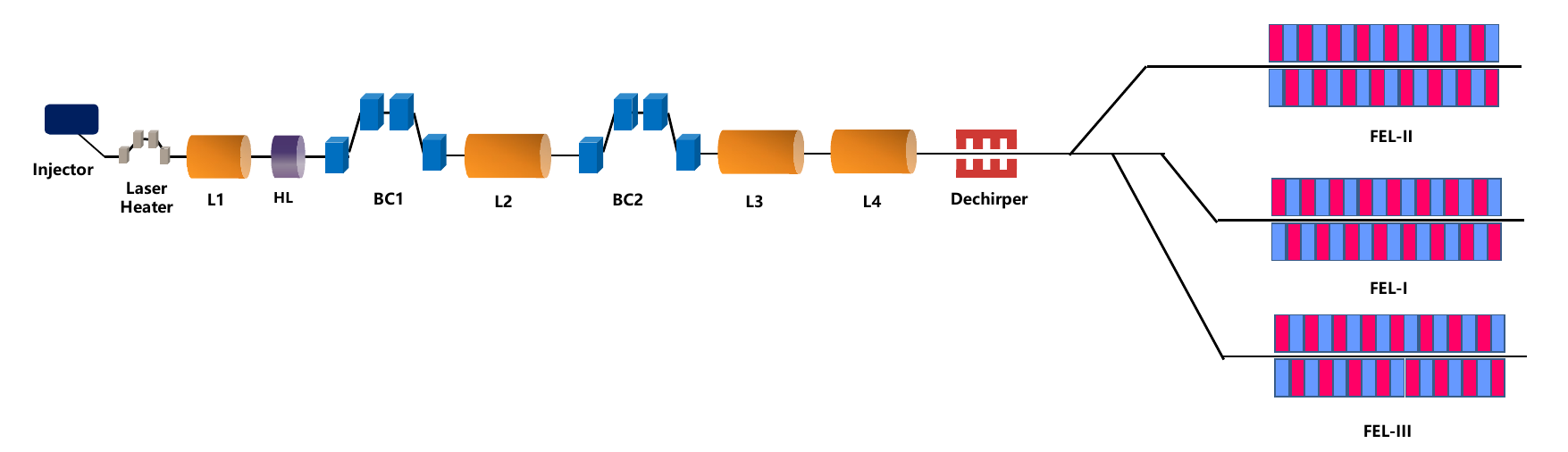}
	\caption{Layout of the SHINE facility.}
	\label{layout}
\end{figure}

In this section, the beam dynamics simulation based on the SHINE design is presented using the proposed scheme. The facility layout is presented in Fig.~\ref{layout}. The photoinjector section of SHINE  provides the electron bunch charge of 100 pC. The 216.7 MHz VHF gun cavity is followed by the 1.3 GHz buncher modifying the bunch longitudinal distribution with velocity bunching. The injector also consists of a single 9-cell cryomodule and an eight 9-cell cryomodule to accelerate the electron bunch to 100 MeV. A laser heater is used to suppress the microbunching instability. The main accelerating section includes the superconducting linac L1, which runs at the off-crest phase to produce the energy chirp for bunch compression. At the end of L1 is the 3.9 GHz harmonic module to linearize the longitudinal phase space. The first bunch compressor chicane BC1 is located behind L1 at a beam energy of nearly 300~MeV. The second accelerating section, L2,  accelerates the beam to a beam energy of around 2.05 GeV and provides the necessary energy chirp for the final longitudinal charge compression, which takes place in BC2. Downstream of BC2 there are two accelerating sections, L3 and L4, to increase the beam energy to 5 and 8 GeV, respectively. A corrugated structure is placed at final energy as a dechirper to compensate for the correlated energy spread.

Downstream of the linac section, the beam is distributed in up to 6 undulators, located in three tunnels. Three additional undulators is foreseen for future upgrades. The modified design with integrated kickers in BC1 and BC2 aims to provide electron bunches with two different longitudinal properties in the linac to drive the FEL lasing in the corresponding configuration. In the previous design of SHINE, the three undulator lines are driven by the same 8-GeV electron bunch to generate FEL pulses with different designed wavelength ranges at repetition rate up to 1 MHz \cite{Liu2023}. As mentioned before, the first and third undulator lines (FEL-I and FEL-III) are operated in SASE \cite{Kondratenko1980,BONIFACIO1984251} and self-seeding mode \cite{FEL1997,Gianluca2011,amann2012,inoue2019,nam2021,liu2023cascaded} to cover the photon energy of 3-15 keV and 10-25 keV, respectively. The FEL-II is specifically designed as a soft X-ray undulator line, covering a photon energy range from 0.4 to 3 keV. It is the most versatile undulator line, incorporating various modes such as SASE, self-seeding, high-gain harmonic generation (HGHG) \cite{Yu1991,Yu2000,Allaria2012}, echo-enabled harmonic generation (EEHG) \cite{Stupakov2009,Xiang2009,Zhao2012,RebernikRibic2019}, and other advanced FEL modes \cite{Yan2021,yan20212,Feng22,yang2022optimization,Yang2023}. The FEL-II could employ the EEHG-HGHG cascading scheme to generate fully coherent soft X-ray FEL pulses. This configuration requires highly stringent spatial and temporal synchronization between the electron bunch and the seed laser, thus necessitating a more uniform longitudinal beam phase space with a relatively longer bunch length \cite{zhu2022inhibition,yang2022optimization}. In contrast, it also requires electron bunches with a high peak current to produce high-power SASE pulses in FEL-I and FEL-III. As a result, stronger longitudinal compression is indispensable to achieve a high peak current of more than 1.5 kA, which corresponds to a bunch length of about 30 fs for a 100 pC bunch. 

Here, we demonstrate the feasibility and effectiveness of the dipole-kicker scheme through start-to-end simulations based on SHINE physical design. The beam dynamics simulations based on SHINE are conducted with the ELEGANT, which includes collective effects such as longitudinal space charge, coherent synchrotron radiation, and wakefield effects \cite{borland2000elegant}. The GENESIS code is used to perform FEL simulations \cite{reiche1999genesis}. If the kicker combination exerts the bump in the same direction as the bending dipoles, the electron bunch will experience stronger longitudinal compression and a slight arrival time delay in the following RF section. Since the kickers can add a flexible bump on the nominal beam trajectory, they can kick the bunches in or against the bending direction to enhance or weaken the compression.  

\begin{figure}[h] 
	\centering 
    \includegraphics*[width=0.9\columnwidth]{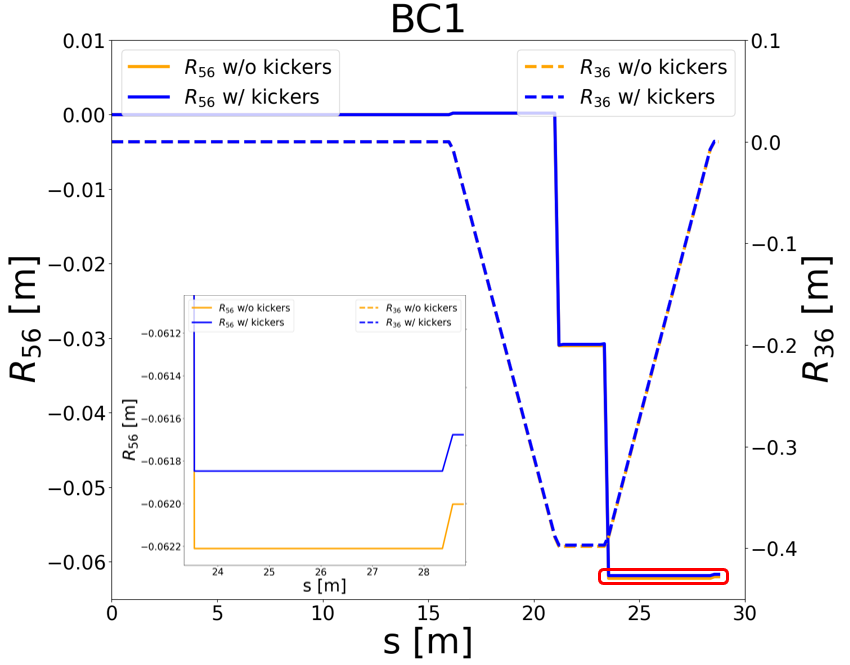}
	\caption[width=1\textwidth]{The evolution of dispersive function values $R_{56}$ and $R_{36}$ along the BC1 from the two mode.}
	\label{matrix_bc1} 
\end{figure}

We consider two electron beam modes, which have the same bunch properties until the entrance of BC1. Here, mode 1 is referred to as the nominal operation without activating the fast kickers, while electron bunches operated in mode 2 are exerted with an additional bump strength from the fast kickers in the dispersive sections. Table~\ref{table1} presents the detailed chicane parameters and beam dynamics properties of the two modes. Note that in SHINE, bunches in the chicane are deflected in the vertical plane. Here, the strength from kickers is decreasing the bending angle of dipole magnets, resulting in a reduced $R_{56}$ strength 
and an early arrival in the L2 section. Though the momentum compaction factor is smaller, it can introduce a larger chirp required in the following dispersive section to achieve the target peak current. Figure.~\ref{matrix_bc1} presents the comparison between the two modes of dispersive function values $R_{56}$ and $R_{36}$ along BC1. 

\begin{table}[h!]
	\begin{center}
		\caption{Parameters of two bunch modes at BC1.}
		\label{table1}
		\begin{tabular}{l l l l}
			\textbf{Parameter}  & \textbf{Unit} & \textbf{Mode 1} & \textbf{Mode 2} \\
			\hline
			Nominal beam energy & MeV& 297& 297\\
			Energy spread & \%&  1.583 & 1.583\\
			Drift length between $1^{st}$ \& $2^{nd}$  & m & 4.7143 & 4.7143\\
			Drift length between $2^{nd}$ \& $3^{rd}$  & m &1.75 & 1.75\\
			Length of dipole magnet & m & 0.2 & 0.2\\
			Length of kicker magnet &  m & 0.2 & 0.2\\
			Dipole Magnetic field & T & 0.39 & 0.39\\
			Kicker Magnetic field &  mT & 0 & -1.10\\
			Bending angle  & degree & 4.526 & 4.539\\
			Momentum compaction  $R_{56}$ & mm & -62.003  &-61.677\\
			Phase difference in L2 &  degree & 0 & -0.265 \\
	         \hline
		\end{tabular}
	\end{center}
\end{table}

\begin{figure}[h] 
	\centering 
	\subfigure[]{
		\label{bc1lps1}	
		\includegraphics[width=0.48\linewidth]{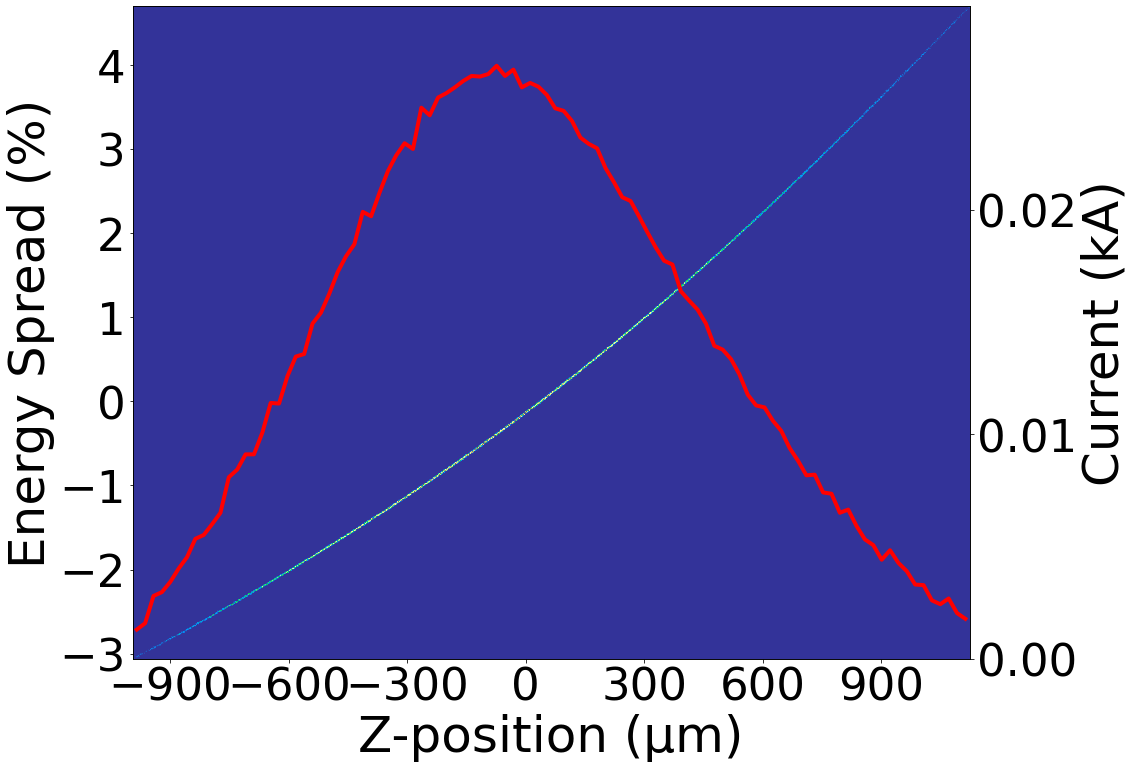}}
	\subfigure[]{
		\label{bc1lps2}	
		\includegraphics[width=0.48\linewidth]{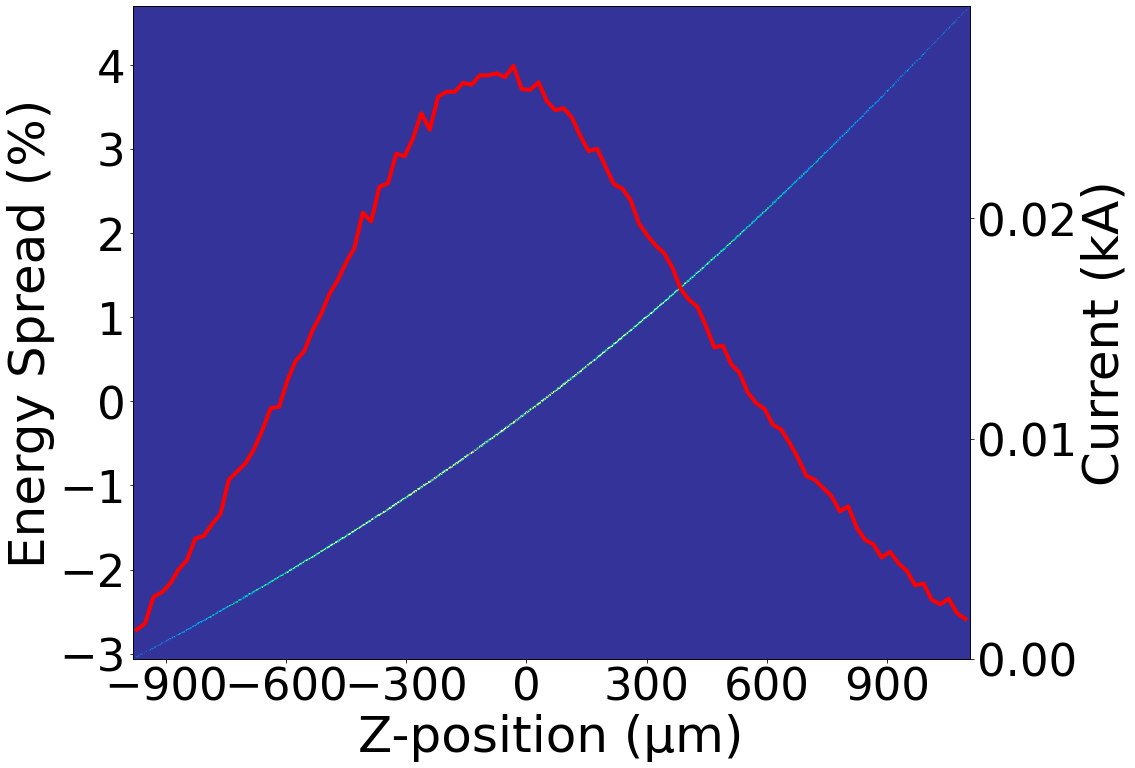}}
	\caption[width=1\textwidth]{Longitudinal phase space distributions at the BC1 end of two modes. (a) for mode 1, (b) for mode 2.}
	\label{bc1lps} 
\end{figure}

The longitudinal phase space of the beam is presented at the BC1 exit for the two operation modes in Fig.~\ref{bc1lps}. No significant difference can be found between the two operational modes. This is attributed to the fact that the beam energy and chirp upon entering the bunch compressor of both modes are identical. Only a minor difference in density modulation introduced by the kicker magnets exists, which causes slight deviations in the trajectory of the beam transport within the dispersive section. This introduced variation in the arrival times can be estimated by Eq.~\ref{eq5}. As the L2 section imprints the required energy chirp on the beam longitudinal phase space using an off-crest accelerating phase, this scenario varies the energy chirps of bunches from the two operational modes. 

For bunches operated in mode 2, the $R_{56}$ strength of the chicane is reduced, but its accelerating phase in the cryomodules will deviate away from the crest, increasing the introduced energy chirp. Meanwhile, due to the reduced energy gain, the bending angles imposed by the dipole magnets in BC2 will be larger. Moreover, the kicker combination adds an additional kick strength to enhance the compression in BC2. These factors can work together to adjust the compression in the second-stage bunch compressor to induce the required final compression. The chicane parameters and beam properties at the second dispersion section are shown in Table~\ref{table2}.

\begin{table}[h!]
	\begin{center}
		\caption{Parameters of two bunch modes at BC2.}
		\label{table2}
		\begin{tabular}{l l l l}
			\textbf{Parameter}  & \textbf{Unit} & \textbf{Mode 1} & \textbf{Mode 2} \\
			\hline
			Nominal beam energy & MeV& 2056.5 & 2050.6\\
			Energy spread & \%&  1.584 & 1.584\\
			Drift length between $1^{st}$ \& $2^{nd}$  & m & 9.9072 & 9.9072\\
			Drift length between $2^{nd}$ \& $3^{rd}$  & m &1.75 & 1.75\\
			Length of dipole magnet &m& 0.55 & 0.55\\
			Length of kicker magnet &  m & 0.2 & 0.2\\
			Dipole Magnetic field & T & 0.567 & 0.567\\
			Kicker Magnetic field &  mT & 0 & 3\\
			Bending angle  & degree & 2.616 & 2.621\\
			Momentum compaction $R_{56}$ & mm &-43.950  & -44.290\\
			Phase difference in L3 &  degree & 0 & 0.2808 \\
	         \hline
		\end{tabular}
	\end{center}
\end{table}

 \begin{figure}[h] 
	\centering 
    \includegraphics*[width=0.9\columnwidth]{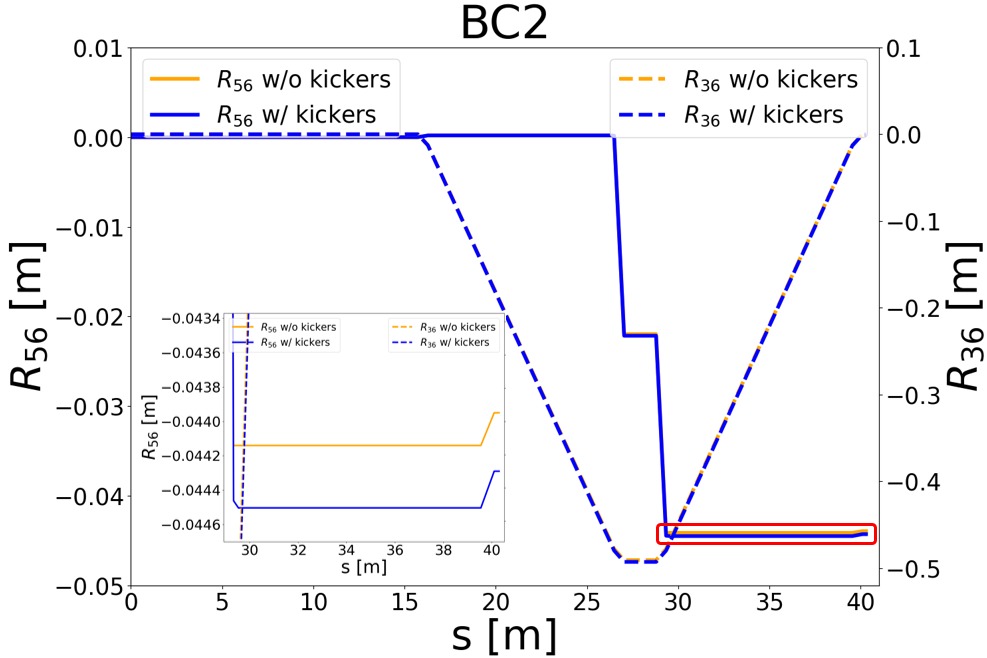}
	\caption[width=1\textwidth]{The evolution of dispersive function values $R_{56}$ and $R_{36}$ along the BC2 from the two mode.}
	\label{matrix_bc2} 
\end{figure}

\begin{figure}[h] 
	\centering 
	\subfigure[]{
		\label{bc2lps900}	
		\includegraphics[width=0.7\linewidth]{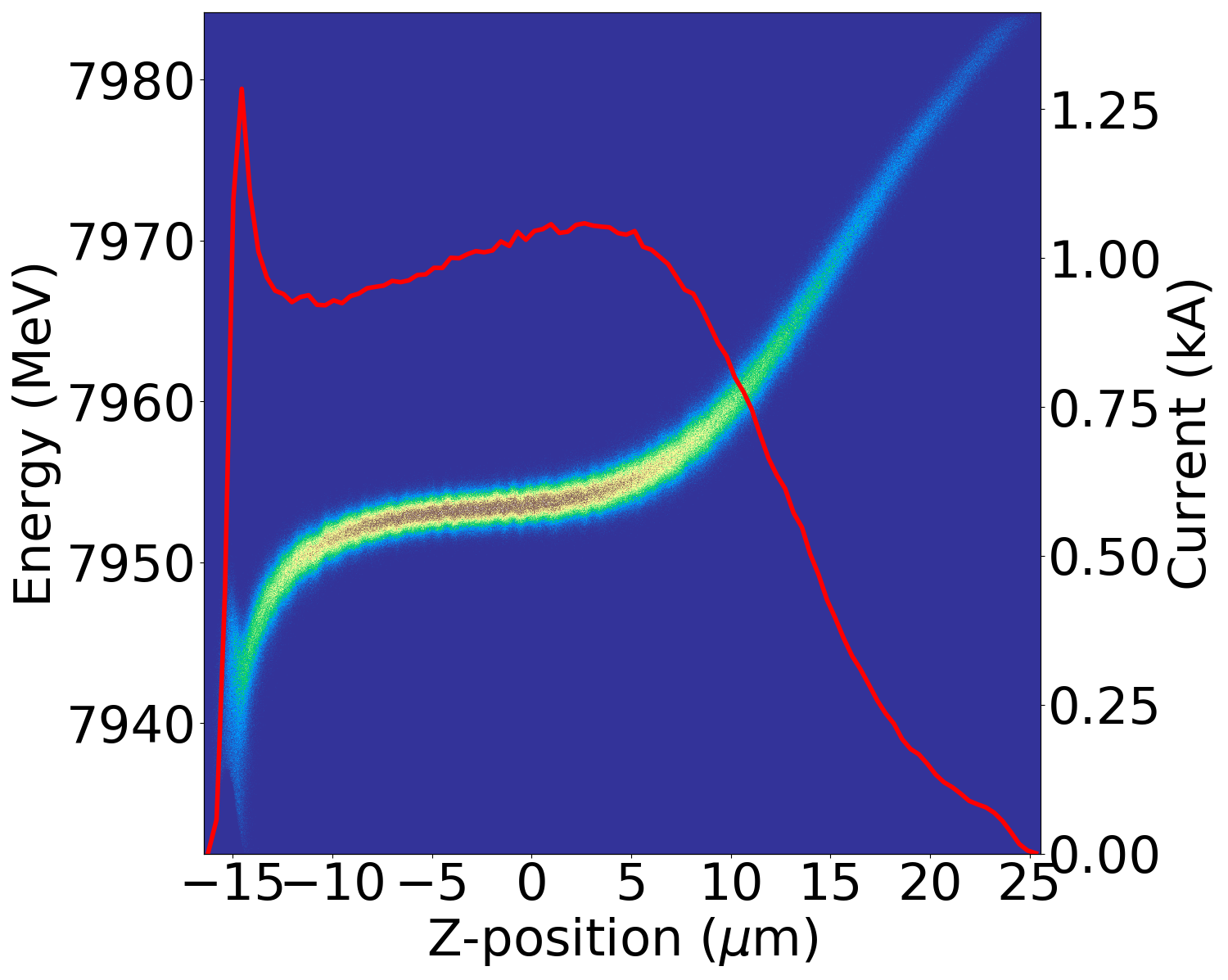}}
	\subfigure[]{
		\label{bc2lps1500}	
		\includegraphics[width=0.66\linewidth]{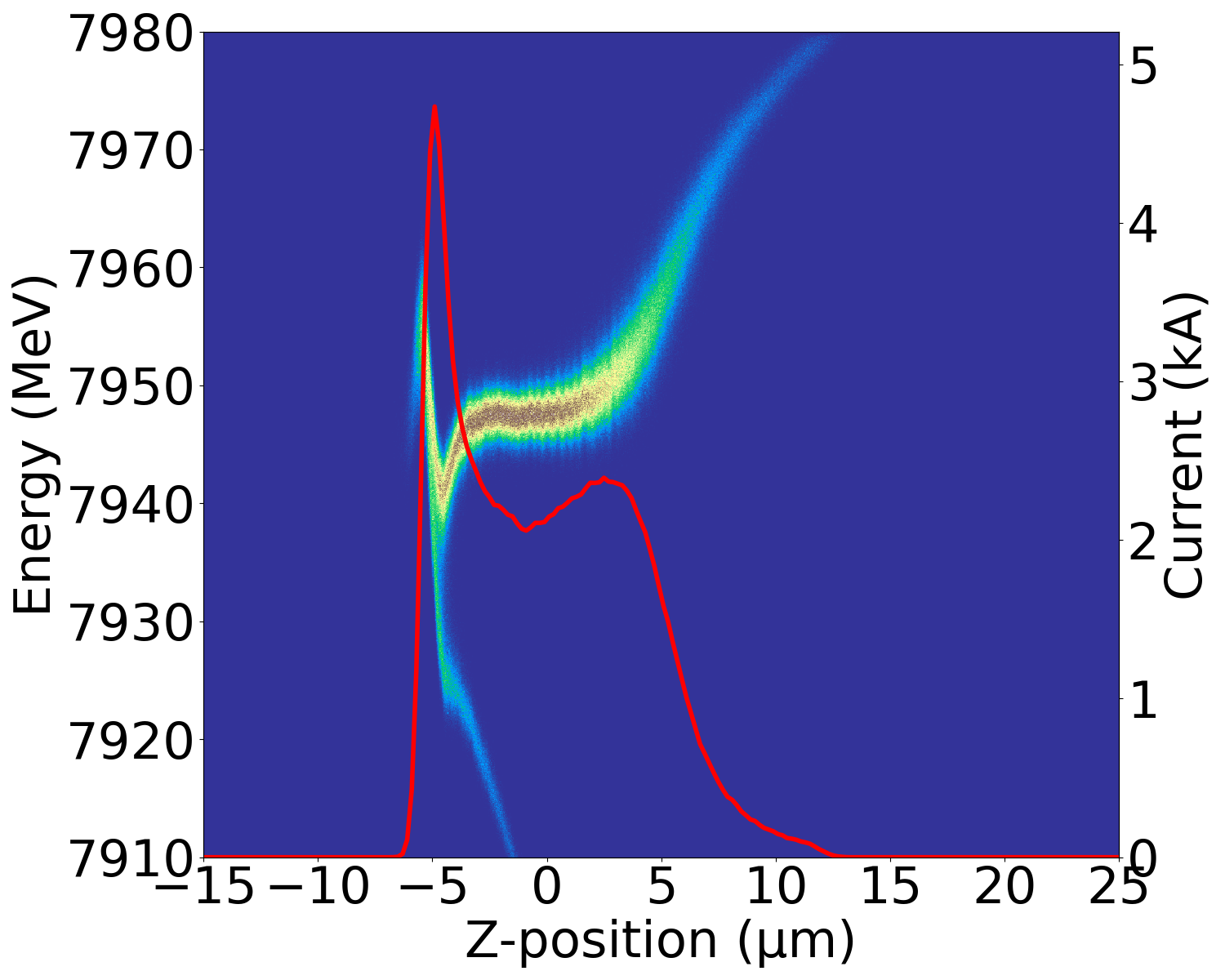}}
	\caption[width=1\textwidth]{Longitudinal phase space distributions at the linac end of two modes. (a) for mode 1, (b) for mode 2.}
	\label{bc2lps} 
\end{figure}

In the SHINE linac design, the compression in BC2 is much larger than in BC1. The more the modulation, the more distortion can be observed in the final longitudinal phase space. Fig.~\ref{matrix_bc2} shows the comparison between the two modes of dispersive function values $R_{56}$ and $R_{36}$ along the BC2 section, from which we can find the maximum dispersion in the chicane sections is around 0.5 m. The final longitudinal phase space distributions are presented in Fig.~\ref{bc2lps}. The RMS bunch length can vary from 3.9 $\rm {\mu m}$ to 10.7 $\rm {\mu m}$ with a slight deviation of beam energy from the nominated mode by about 5 MeV at a nominal value of around 8 GeV. It demonstrates that the proposed scheme successfully realizes the flexible bunch length control in a CW XFEL facility. 

\begin{figure}[h] 
	\centering 
	\subfigure[]{
		\label{trans900}	
		\includegraphics[width=0.8\linewidth]{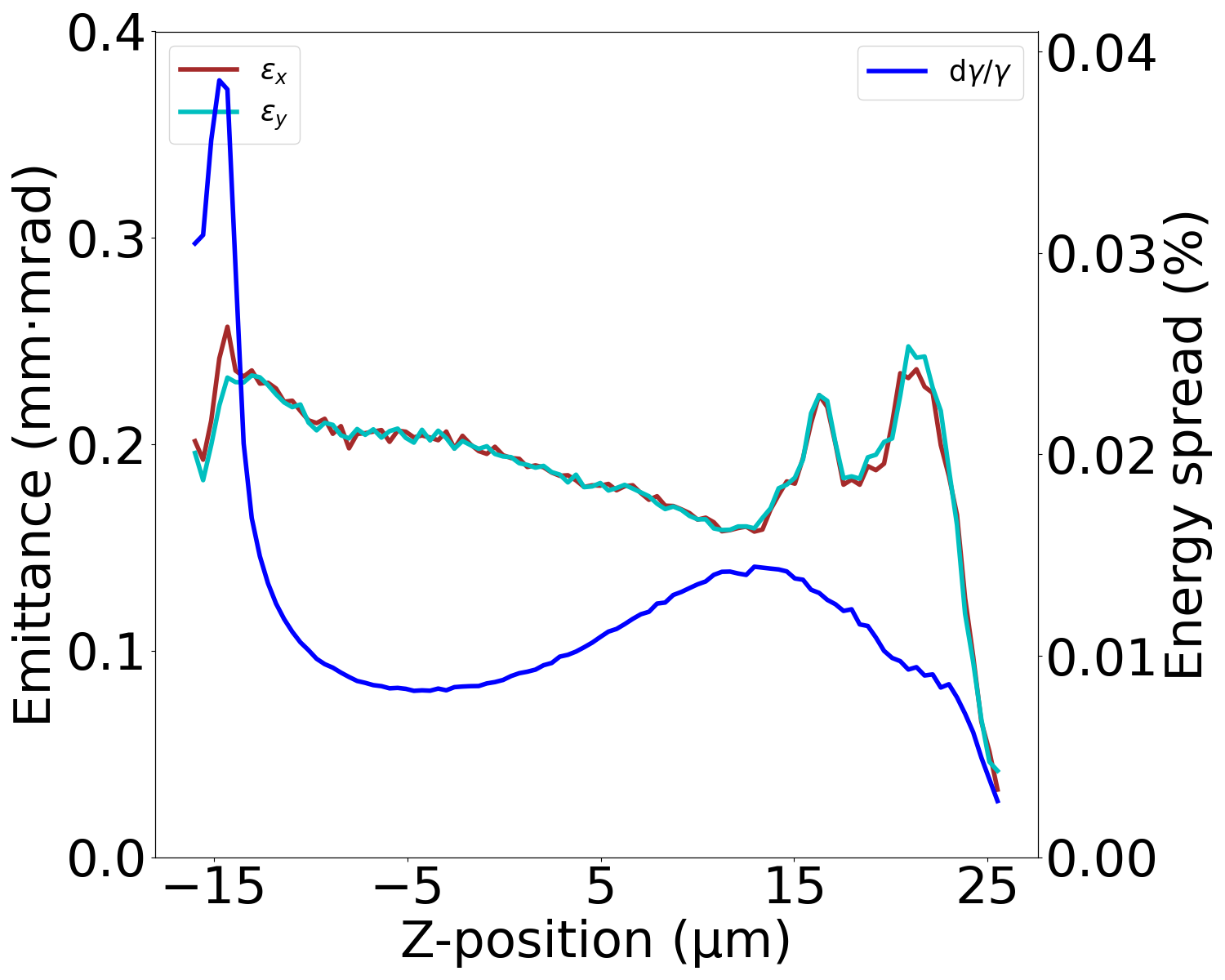}}
	\subfigure[]{
		\label{trans1500}	
		\includegraphics[width=0.8\linewidth]{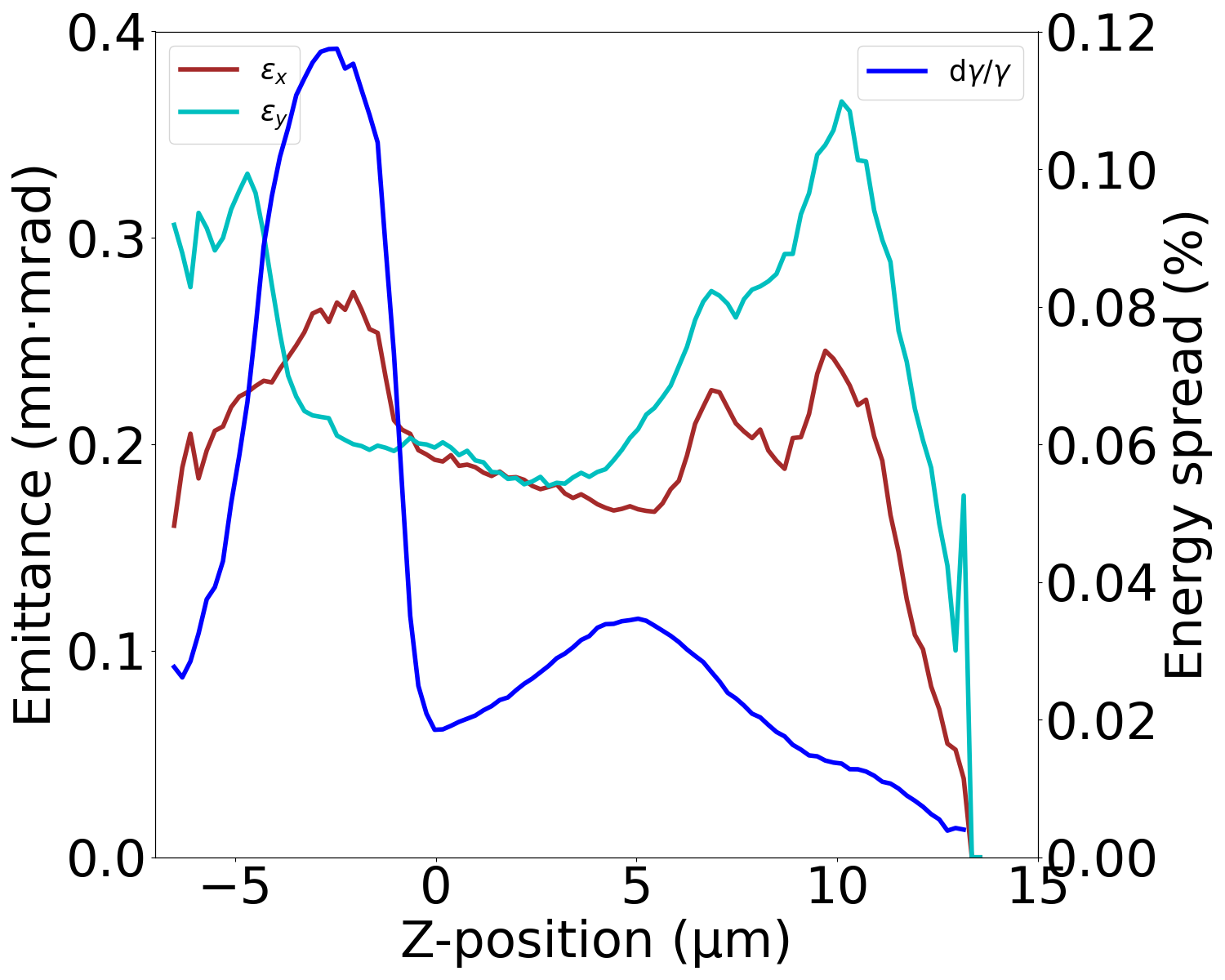}}
	\caption[width=1\textwidth]{Beam slice properties of the two operation modes. (a) for mode 1, (b) for mode 2.}
	\label{bc2tran} 
\end{figure}

Furthermore, Fig.~\ref{bc2lps1500} shows the typical single-spike shape in the current profile after undergoing intense nonlinear density modulation, which is distorted by nonlinear effects. A dense distribution at the head and/or tail of the bunch is a common phenomenon due to the non-uniform compression ratios along the longitudinal distribution. This amplifies slice energy spread and slice transverse emittance. Figure~\ref{bc2tran} compares slice beam properties between the two operation modes. It illustrates that for short bunch mode with high peak current, the favorable slice parameters are well-preserved in the bunch core, meeting the requirements for subsequent SASE mode operation. Thus, the kickers in the bunch compressor accomplish the final bunch length adjustment by more than a factor of two in the core of the bunch, allowing a change of operation mode from externally seeded FEL to SASE FEL. 

Because the beam in the cryomodule downstream of the BC1 has a different accelerating phase with the kickers switched on, there is a small deviation of beam energy at the entrance of the second bunch compressor. As the bunches from mode 1 and mode 2 are inverted in the two bunch compressors (decreased $R_{56}$ strength in BC1, but increased $R_{56}$ strength in BC2), the final bunch energy from the two modes doesn't differ much. The final energy of the long bunch is 7955 MeV and 7949 MeV for the short one, as can be examined in Fig.~\ref{bc2lps}.

\begin{figure}[h] 
	\centering 
	\subfigure[]{
		\label{fel2_power1}	
		\includegraphics[width=0.45\linewidth]{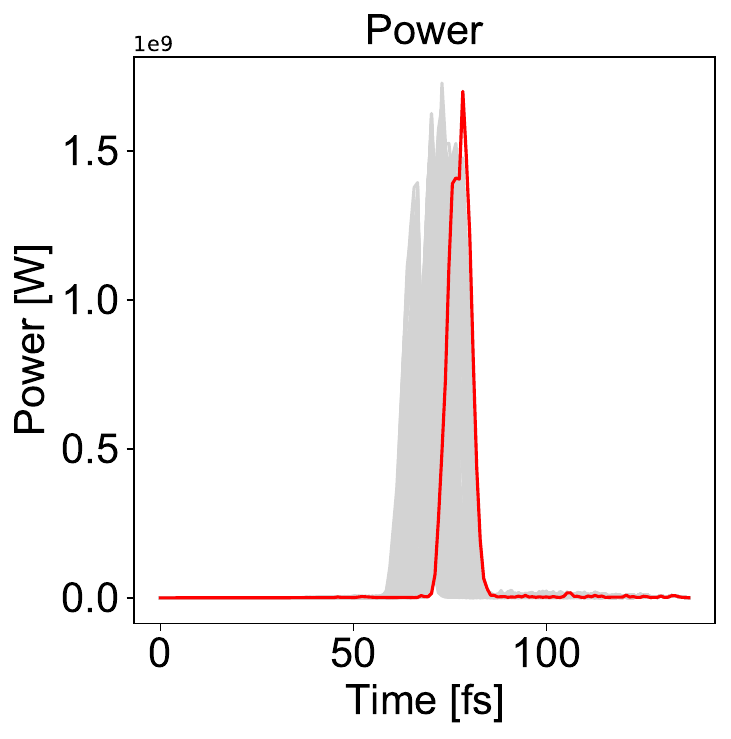}}
	\subfigure[]{
		\label{fel2_spectrum1}	
		\includegraphics[width=0.46\linewidth]{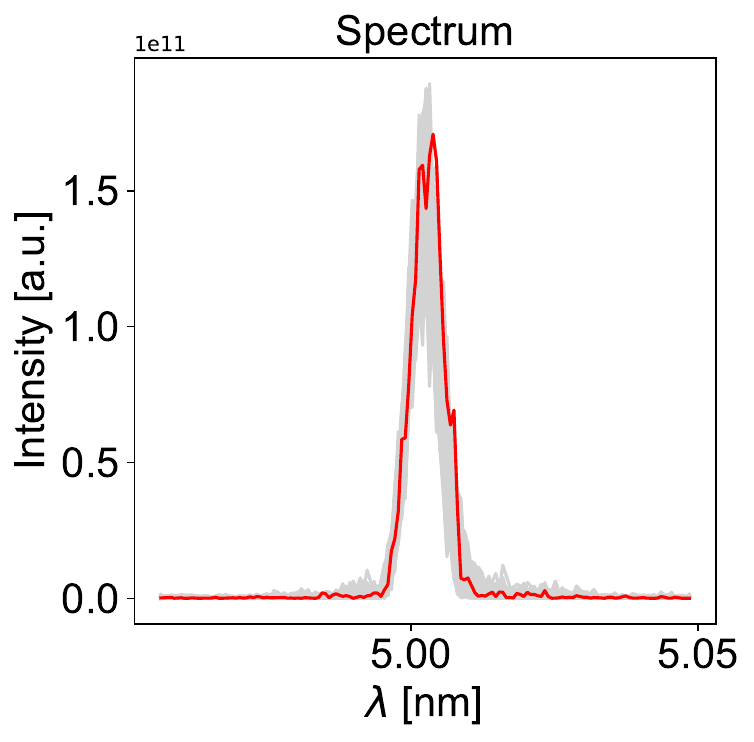}}
	\caption[width=1\textwidth]{The 100 shots output FEL performance from the first stage EEHG. The gray lines represent FEL pulses with different temporal interaction positions between the electron beam and the seed laser. The red line represents one of the shots.}
	\label{fel2_result1} 
\end{figure}
\begin{figure}[h] 
	\centering 
	\subfigure[]{
		\label{fel2_power2}	
		\includegraphics[width=0.45\linewidth]{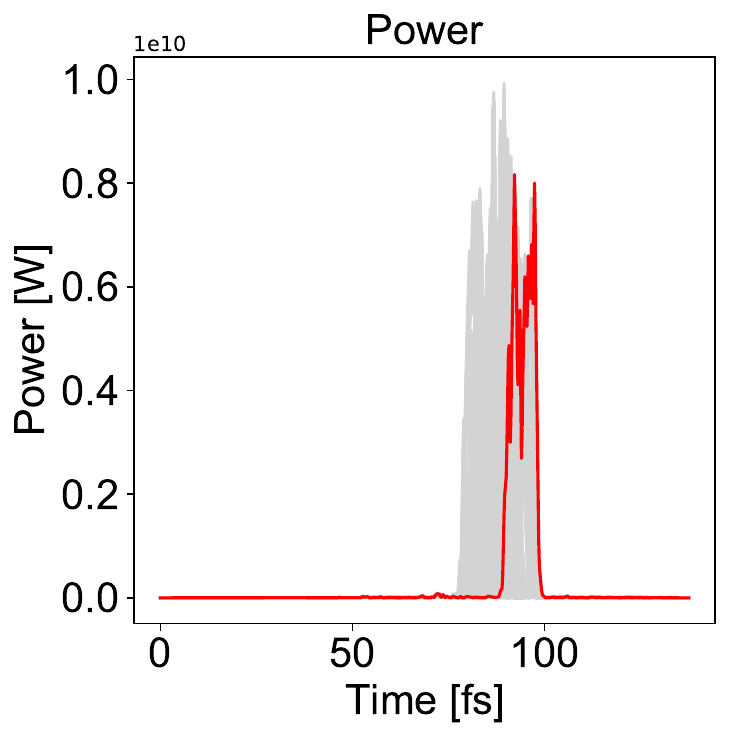}}
	\subfigure[]{
		\label{fel2_spectrum2}	
		\includegraphics[width=0.475\linewidth]{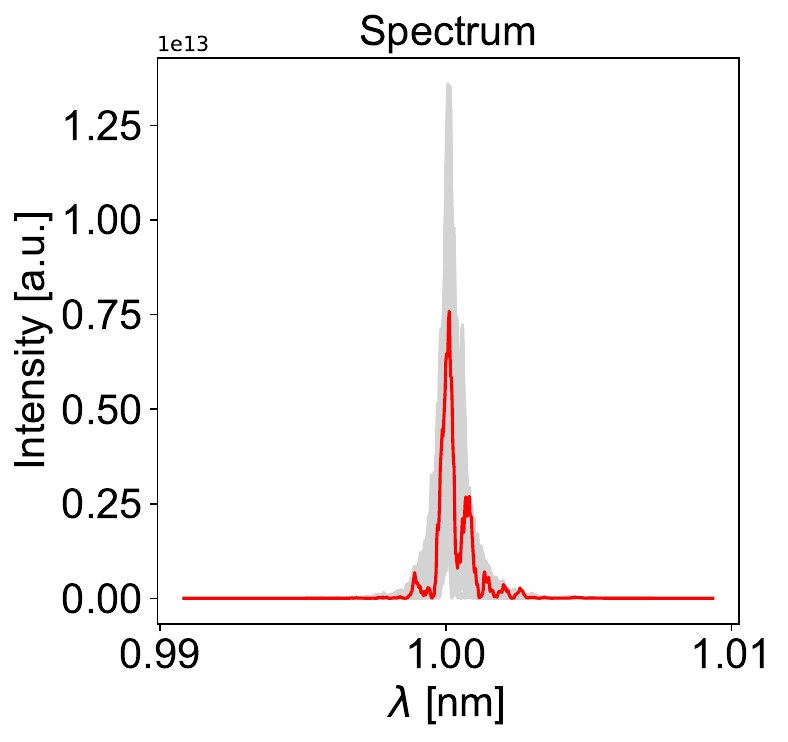}}
	\caption[width=1\textwidth]{The 100 shots final output FEL performance from the second stage HGHG. The gray lines represent FEL pulses with different temporal interaction positions between the electron beam and the seed laser. The red line represents one of the shots.}
	\label{fel2_result2} 
\end{figure}

To demonstrate the electron beam parameters are satisfied for both the externally seeded FEL and the SASE FEL, we simulate the FEL performance in both FEL-II and FEL-III undulator lines, respectively. The longer bunches of mode 1, with the kickers switched off, are deflected to the FEL-II which is designed to the cascaded EEHG-HGHG configuration \cite{yang2022optimization}. In our simulations, two seed lasers can be split precisely from optics, resulting in negligible relative timing jitter. For the sake of simplicity, we only consider the relative timing jitter of the seed laser and the electron beam at the entrance of the modulators. The pulse duration of the seed lasers is 20 fs (FWHM), and a relative timing jitter of 3 fs (RMS) is assumed. Figure~\ref{fel2_result1} presents the power and spectrum of the output FEL performance after the first stage. The mean pulse energy of the 5 nm FEL after the first stage EEHG is 10.24 $\rm \mu$J with a standard deviation of 0.7 $\mu$J and a jitter of 6.87~\%. The output FEL from the first stage EEHG is used to seed the second stage HGHG with the fresh bunch technique. Figure.~\ref{fel2_result2} shows the final output FEL performance after the second stage HGHG at 24 m, with a mean pulse energy of 46.43 $\mu$J, a standard deviation of 2.29 $\mu$J, and a jitter of 4.93~\%. Driven by the long electron bunches of mode 1, the output FEL pulse energy stability of the cascaded EEHG-HGHG configuration is at least four times better than that of the short-bunch operation. 

\begin{figure}[h!] 
	\centering 
	\subfigure[]{
		\label{fel3_power}	
		\includegraphics[width=0.48\linewidth]{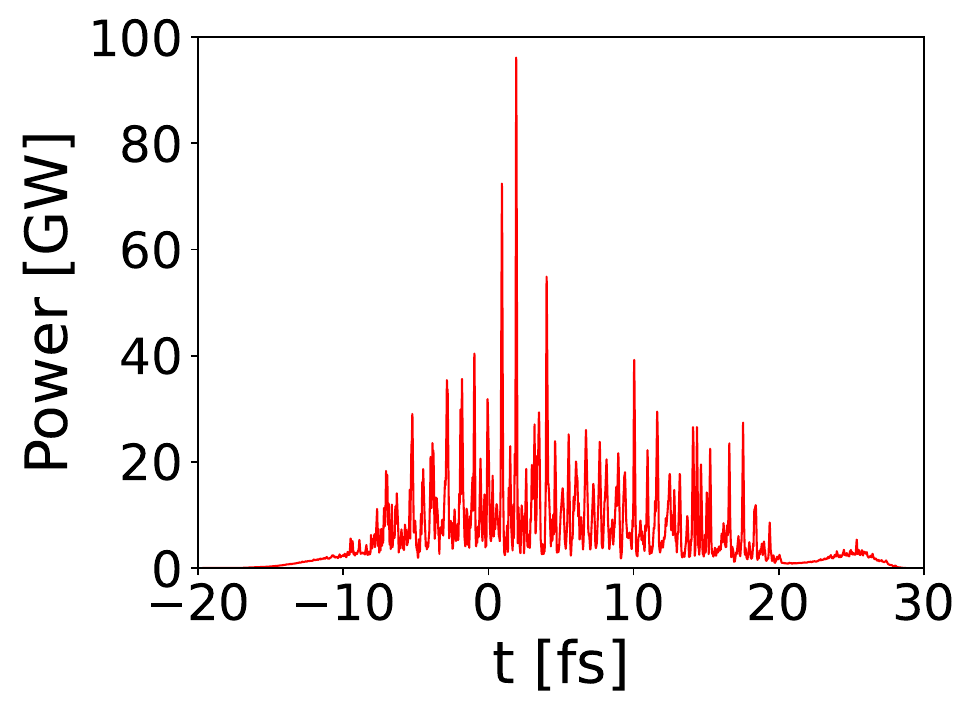}}
	\subfigure[]{
		\label{fel3_spectrum}	
		\includegraphics[width=0.48\linewidth]{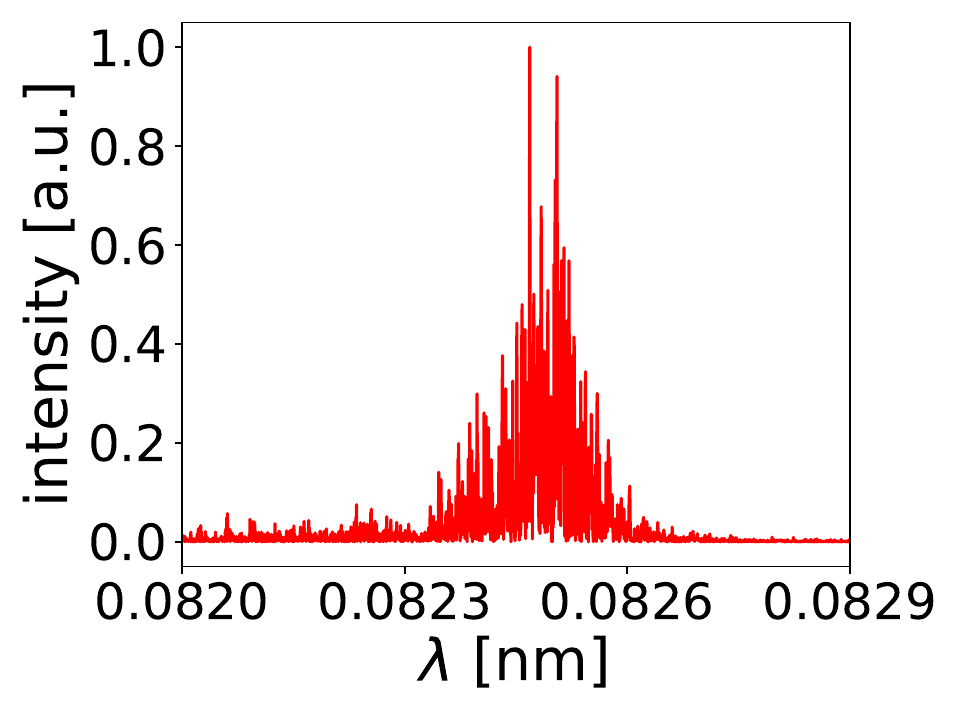}}
	\caption[width=1\textwidth]{Simulated power profile (a) and spectrum (b) of the SASE pulse using the electron beam of the short-bunch mode.}
	\label{fel3_result} 
\end{figure}

The baseline operation mode of FEL-III is SASE, which prefers shorter bunch with a higher peak current. The simulated FEL performance at the photon energy of 15 keV based on the electron bunch of mode 2 is presented in Fig.~\ref{fel3_result}. Despite the non-uniform compression leading to spike formations at the head and tail of the bunch, the transverse emittance and slice energy spread in the central part of the bunch are still well-preserved. These properties ensure a good SASE performance, resulting in a pulse energy of 291.54 $\rm \mu$J and a peak power of 96 GW.

\section{Discussion}
\label{s5}
We verify the feasibility of the dipole-kicker combination with start-to-end simulations. While these simulations demonstrate the potential of the proposed scheme, additional technical considerations must be addressed before implementing it in practical operation at a continuous wave (CW) XFEL user facility.

One significant question is the availability of kick magnets in terms of switching speed, stability, and strength. The resonant kickers installed at SwissFEL have the required properties. However, it is unclear regarding the rapidity with which both flanks go to zero before the next bunch arrives. The kickers installed for the intermediate dump at European XFEL have the speed and amplitude of the magnetic field, but the stability remains a concern. Hence the feasibility of extending this system to CW machines requires more exploration. 

One approach to mitigating instability is to power all the kickers in series with a single power supply. With stability of the order of one percent, this configuration will not result in a (fluctuating) transverse angle and offset and the jitter will only result in a change in $R_{56}$. However, the effect is of percent level of an already small effect and is believed to play a minor role. This should nevertheless be checked in a future study. With the kickers in series, the amplitude of the kicker pulses will attenuate with each kicker. The difference in travel time between electron bunches and field between the kickers gives rise to a shift in time along the kicker pulse from kicker to kicker, which in turn leads to a minimum length of the kicker pulse required or, for a given kicker, on the maximum length of the chicane in which they are installed. The loss in kicker power can be compensated by adding a chirp on the kicker pulse which exactly compensates the attenuation.

Another potential problem could arise if quadrupoles are integrated into the chicane. As the trajectory within the chicane is different with kickers powered on and off, these kinds of chicanes can only be used if the quadrupoles are outside of the kicker system. If the kickers are placed between the outermost dipoles, the quadrupoles would need to be in front of the first and behind the last kicker. In the BC2 section of SHINE, it is possible, but the BC1 chicane would need to be much longer than currently designed, otherwise, a scheme with an integrated quadrupole can not be implemented.

\section{Conclusion and Outlook}
\label{s6}
This paper proposes a multi-bunch-length operation for CW XFEL facilities based on fast kickers. This method facilitates the generation of bunch patterns with arbitrary compression scenarios to drive the FEL lasing in several undulator lines based on each specific operation mode. Start-to-end simulations based on SHINE physics design demonstrate that this dipole-kicker combination scheme can realize bunch-to-bunch current profile control that meets the requirements for different undulator lines. Although only the operation mode of two bunch lengths is presented in this paper, this approach can also be extended to arbitrary bunch length modes. The kicker amplitudes assumed in this paper are rather conservative and were derived from the specifications at European XFEL with very large apertures of around 30~mm. With a reduced kicker aperture, the kicker amplitude can easily be increased by a factor of 2 with the same stability parameters.

\section{Acknowledgments}
This work was supported by the National Natural Science Foundation of China (Nos. 12122514 and 11975300), Shanghai Science and Technology Committee Rising-Star Program (20QA1410100). Bart Faatz was funded by the Chinese Academy of Sciences President’s International Fellowship Initiative (PIFI), Grant No. 2020FSM0003. We thank Frank Obier from DESY for useful discussions on the XFEL kicker system.

\bibliographystyle{elsarticle-num}
\bibliography{ref}

\end{document}